\documentclass[]{ceurart}

\usepackage{cleveref}
\usepackage{subcaption}

\begin{document}

\copyrightyear{2023}
\copyrightclause{Copyright for this paper by its authors. Use permitted under Creative Commons License Attribution 4.0 International (CC BY 4.0).}

\conference{Barcelona'23: 6th Workshop on Natural Language Processing for Requirements Engineering, April 17, 2023, Barcelona, ES}

\title{Let's Stop Building at the Feet of Giants: Recovering unavailable Requirements Quality Artifacts}

\author[1]{Julian Frattini}[email=julian.frattini@bth.se, orcid=0000-0003-3995-6125]
\address[1]{Blekinge Institute of Technology, Valhallavägen 1, 371 41 Karlskrona, Sweden}

\author[2]{Lloyd Montgomery}[email=lloyd.montgomery@uni-hamburg.de, orcid=0000-0002-8249-1418, url=https://lloydm.io/]
\address[2]{University of Hamburg, 20146 Hamburg, Germany}

\author[1]{Davide Fucci}[email=davide.fucci@bth.se, orcid=0000-0002-0679-4361, url=https://dfucci.github.io/]
\author[3,4]{Jannik Fischbach}[email=jannik.fischbach@netlight.com,orcid=0000-0002-4361-6118]
\address[3]{Netlight Consulting GmbH, Sternstraße 5, 80538 Munich, Germany}
\address[4]{fortiss GmbH, Guerickestraße 25, 80805 Munich, Germany}

\author[1]{Michael Unterkalmsteiner}[email=michael.unterkalmsteiner@bth.se, orcid=0000-0003-4118-0952, url=https://www.lmsteiner.com/]
\author[1,4]{Daniel Mendez}[email=daniel.mendez@bth.se, orcid=0000-0003-0619-6027, url=https://www.mendezfe.org/]

\begin{abstract}
  Requirements quality literature abounds with publications presenting artifacts, such as data sets and tools. However, recent systematic studies show that more than 80\% of these artifacts have become unavailable or were never made public, limiting reproducibility and reusability. In this work, we report on an attempt to recover those artifacts. To that end, we requested corresponding authors of unavailable artifacts to recover and disclose them according to open science principles. Our results, based on 19 answers from 35 authors (54\% response rate), include an assessment of the availability of requirements quality artifacts and a breakdown of authors' reasons for their continued unavailability. Overall, we improved the availability of seven data sets and seven implementations.
\end{abstract}


\maketitle

\section{Introduction}
\label{sec:introduction}

Data sets and tools are often reported as important contributions to requirements quality literature~\cite{montgomery2022empirical}. However, a recent secondary study revealed that out of 57 primary studies, as little as 12\% of data sets and 19\% of tools are currently publicly available~\cite{frattini2022live}. The unavailability of those artifacts has two major consequences. Firstly, empirical results are difficult to \textit{reproduce}, which inhibits the process of strengthening the empirical evidence of scientific contributions. Secondly, the presented artifacts are difficult to \textit{reuse}, which necessitates scientific progress to restart over and over again instead of evolving from existing contributions. Ultimately, these consequences inhibit the progress of the requirements quality research domain.

Following open science practices in software engineering improves the accessibility of artifacts and the preservation of future contributions~\cite{mendez2020open} but it is difficult to apply them in retrospect. Because of this, the accessibility of artifacts in past publications deteriorated over time~\cite{montgomery2022empirical}. The resulting unavailability of artifacts~\cite{frattini2022live} poses a significant challenge to artifact-dependent research, like the requirements quality or the larger natural language processing for requirements engineering (NLP4RE) domain. In this work, we set to recover unavailable requirements quality artifacts by requesting authors to disclose them according to open science principles. Our main contribution is the recovery of seven data sets and seven implementations. Nevertheless, 16 out of 35 (46\%) requests to authors remained unanswered, indicating that the research community needs to emphasize further the importance of persistently archiving research data.

The remainder of this manuscript is structured as follows: \Cref{sec:background} introduces the background both on the topic of open science and requirements quality research. \Cref{sec:recovery} describes the process and \Cref{sec:results} the results of the recovery. We discuss these results in \Cref{sec:discussion} before concluding in \Cref{sec:conclusion}.

\section{Background}
\label{sec:background}

\subsection{Open Science in Software Engineering}

Scientific work needs to be reproducible~\cite{mendez2020open} to strengthen the evidence it contributes to a field of research~\cite{anda2008variability}. \textit{Open science} is the initiative of ensuring public availability of research artifacts~\cite{mendez2020open} and, hence, facilitating reproducibility. Within open science, the facets of \textit{open access} for publications, \textit{open data} for data sets, and \textit{open source} for source code are most relevant to software engineering~\cite{tennant2019foundations}, where each facet of open science entails different techniques and best practices to disclose its respective type of research artifact. Several governmental research funding agencies, including the European Union, made open access to scientific results (including data, tools, etc.) mandatory\footnote{\url{https://research-and-innovation.ec.europa.eu/strategy/strategy-2020-2024/our-digital-future/open-science/open-access_en}}. 


In literature, Minocher et al. propose four attributes \textit{data recoveribility}, \textit{data usability}, \textit{analytical clarity}, and \textit{agreement of results} and explicitly emphasize the sequential dependency of those attributes---e.g., \textit{analytical clarity} of data is meaningless if the data is not \textit{recoverable}. 

Recent endeavors of incentivizing scholars to follow open science principles include open science badges~\cite{kidwell2016badges} and the registered reports~\cite{nosek2018preregistration}. However, the software engineering research community is still in the process of adapting open science principles~\cite{mendez2020open}, and the unavailability of artifacts is common~\cite{montgomery2022empirical}. Prominent reasons for the unavailability of artifacts include the sensitivity of data or corresponding authors changing their affiliation and consequently losing access to their artifacts~\cite{gabelica2022many}. While some reasons for the unavailability of artifacts (e.g., the sensitivity of company-owned data) may well require significant effort to cope with, other reasons (e.g., loss of artifact, lack of diligence) can be circumvented easily by following proposed guidelines~\cite{mendez2020open} and making use of modern tools for artifact sharing.

\subsection{Requirements Quality Literature}

Recent advances have established that artifacts produced in requirements engineering (RE) have a significant impact on downstream software development activities~\cite{wagner2019status}, potentially even causing project failure~\cite{fernandez2013naming}. Consequently, requirements artifacts merit quality assurance~\cite{montgomery2022empirical}. The requirements quality literature is dedicated to providing the understanding as well as the support for measuring and improving the quality of requirements~\cite{montgomery2022empirical}. One popular approach to this is the proposal of \textit{quality factors}. Requirements quality publications often formulate one or more quality factors---e.g., the use of \textit{coordination ambiguity} leading to divergent interpretations~\cite{ezzini2021using}---annotate instances of that quality factor in a \textit{data set}, and finally present an \textit{implementation} (i.e., an algorithm or full-fledged tool) to detect these instances automatically.

These artifacts---both data sets and implementations---represent essential contributions facilitating empirical research and technology transfer. While the (annotated) data sets are the main driver for developing new and improving existing implementations for quality factor detection, implementations are the tools to be deployed in industry for actual integration and improvement of the software engineering process. The NLP4RE research domain, which applies natural language processing (NLP) techniques to RE~\cite{zhao2021natural} and constitutes a large part of the contributions to the requirements quality literature~\cite{montgomery2022empirical}, is particularly focused on said delivery and improvement of tools. In addition to the dependency of these NLP-powered tools on the availability and reliability of training data, this puts the NLP4RE research domain on the forefront of the open science challenge~\cite{zhao2021natural}. The NLP4RE community is therefore particularly aware of its dependency on the availability of artifacts~\cite{dalpiaz2018natural}.

However, recent systematic studies revealed that a significant amount of these artifacts are not available\footnote{Where available means a status of \textit{Upon request} (see \Cref{tab:statuscodes}) or better.} anymore or have never been~\cite{zhao2021natural,montgomery2022empirical,frattini2022live}. \Cref{tab:statuscodes} reports the availability status of 57 data sets (D) and 36 implementations (I) extracted from the 57 primary studies of our previously-published literature review on requirements quality factors~\cite{frattini2022live}. 

\begin{table}[h]
    \centering
    \caption{Availability status of requirements quality artifacts~\cite{frattini2022live} including data sets (D) and implementations (I)}\label{tab:statuscodes}
    \begin{tabular}{l|p{9.5cm}|rr} \toprule
        \textbf{Status} & \textbf{Explanation} & (D) & (I) \\ \hline
        Open Data & The artifact is hosted in a service that satisfies the following criteria: (1) immutable URL (cannot be altered by the author or someone else), (2) permanent (the hosting organization has a mission to maintain artifacts for the foreseeable future), (3) accessible (there is a DOI pointing to the real data source URL), and (4) open-source license (the artifact has a license which grants access and re-use) & 1 & 0 \\
        Open Source\footnotemark & [only for implementations] The implementation is available for all to use, and the code base has been disclosed & - & 5 \\
        Available in paper & [only for data sets] The data set is small enough that the authors disclose the entire data set in the manuscript & 5 & - \\
        Reachable link & The artifact is reachable now but is missing some of the Open Data aspects (see above) & 1 & 1 \\
        Upon request & Authors claim the artifact is available upon request & 0 & 1 \\
        Broken link & A link to the artifact is contained in the paper, but it does not resolve & 10 & 1 \\
        No link & An artifact is presented, but no indication on how to access it is provided & 15 & 27 \\
        Private & The authors state that an artifact exists but is private for some reasons (such as industry collaboration with private data, etc.) & 24 & 0 \\
        Proprietary & The artifact is proprietary, and access is granted upon payment & 1 & 1 \\ \hline
        Total & & 57 & 36 \\
        \bottomrule
    \end{tabular}
\end{table}

\footnotetext[3]{In this context, we are using the term \textit{open source} as commonly understood, not as used in the open science framework~\cite{tennant2019foundations}, which would imply adherence to the properties listed under \textit{open data}.}

\section{Recovery Process}
\label{sec:recovery}

The insight that the availability of requirements quality artifacts is insufficient~\cite{montgomery2022empirical,frattini2022live} motivated our objective to improve the state of open science in the requirements quality literature by ensuring the \textit{recoverability of data}, a necessary prerequisite for the reproducibility of scientific work~\cite{minocher2020reproducibility}.

In this section, we document the artifact recovery process along the undertaken steps. In \Cref{sec:recovery:sample}, we describe the selection of the sample of primary studies. We detail our approach to contact corresponding authors in \Cref{sec:recovery:mails} and maintain correspondence with them in \Cref{sec:recovery:correspondence}. Finally, we document the evaluation of the recovery process and success in \Cref{sec:recovery:evaluation}. All produced data, scripts, and documentation are disclosed in our replication package\footnote{Available at \url{https://doi.org/10.5281/zenodo.7708571}}.

\subsection{Study sample selection and preparation}
\label{sec:recovery:sample}

We used \textit{convenience sampling} since the primary studies on which we base our results were selected based on expediency~\cite{baltes2022sampling}. In particular, we recovered artifacts from a set of primary studies used to build an ontology of requirements quality factors~\cite{frattini2022live}. To develop this ontology, we collected manuscripts reporting \textit{quality factors} from an original set of publications reported in another secondary study~\cite{montgomery2022empirical}. Extracting data sets and implementations from such publications revealed the unfortunate state of artifact availability.

We enhanced the data regarding data sets and implementations from our previous study~\cite{frattini2022live} with the following information.
\begin{itemize}
    \item Corresponding author: each artifact was associated with a corresponding author.
    \item Mention: each artifact was associated with its verbatim mention in the manuscript.
\end{itemize}
Additionally, we corrected information about one data set and three implementations that persisted in the previous study~\cite{frattini2022live}.

Using a spreadsheet, we collected data about
\begin{enumerate}
    \item \textbf{authors} (n=35), specifying for each author the name and email address,
    \item \textbf{data sets} (n=57), specifying for each data set its containing publication, its verbatim mention, the corresponding author, and its current availability, and
    \item \textbf{implementations} (n=36), specifying for each implementation its containing publication, its verbatim mention, the corresponding author, and its current availability.
\end{enumerate}

\subsection{Approaching authors}
\label{sec:recovery:mails}

We created a Python script that automatically assembles one email for each corresponding author. This email contained the following elements:
\begin{enumerate}
    \item Header: an explanation of our endeavor and a request to contribute to open science (or alternatively explain why this is impossible).
    \item Artifact list: a list of artifacts contained in the publications of the authors that were not open access. 
    \item Instructions: brief \textit{how to} to properly disclose artifacts according to the open science principles as well as the offer to assist them in the process
    \item Contact: a way to reach out to us.
\end{enumerate}

We approached the authors in a first mail on the 30\textsuperscript{th} of November 2022, followed by a reminder on the 13\textsuperscript{th} of December, and a final reminder on the 11\textsuperscript{th} of January 2023. For authors that did not respond to our request until the final reminder, we additionally contacted their co-authors to increase the likelihood of response. We concluded the recovery process on the 8{th} of February 2023, yielding a time frame of 70 days.

\subsection{Correspondence}
\label{sec:recovery:correspondence}

We kept close contact with the authors we approached by responding in a window of 24 hours within workdays. During this process, we clarified concerns and offered our help. We processed and recorded the information contained in the authors' answers in a spreadsheet file. We tracked the \textit{response status} in an additional column, denoting the request as either \textit{undeliverable}, \textit{unanswered}, \textit{answered}, or \textit{completed}. We labeled a recovery request as completed once the corresponding author, for all their artifacts, either improved their availability or explained the inability to recover or disclose them. 

Furthermore, we documented the dates of the first email sent, the first response received, and the completion of the request alongside the number of emails sent by the author in addition to the updated availability status of the artifacts and, eventually, the author's explanation for not taking the recommended actions. Two authors coded these explanations independently and came to an absolute agreement on the types of reasons for non-recovery. When the corresponding author's email address was no longer used, we reached out via personal contacts or social networks like Twitter and LinkedIn.

\subsection{Evaluation}
\label{sec:recovery:evaluation}

To evaluate the artifact recovery process, we generated statistics of the following data from the documentation in our tables.
\begin{enumerate}
    \item Correspondence (author response time and frequency) to evaluate the effort of the recovery process.
    \item Recovery request success (change in artifact availability) to evaluate the success of the recovery process.
    \item Reason for non-recovery (author responses excusing the recovery) to evaluate the reasons inhibiting open access.
\end{enumerate}
We evaluated the data by generating descriptive statistics from our documentation.

\section{Results}
\label{sec:results}

\subsection{Correspondence}

\begin{figure}[h]
    \centering
    \begin{subfigure}{.3\textwidth}
        \centering
        \includegraphics[width=\textwidth]{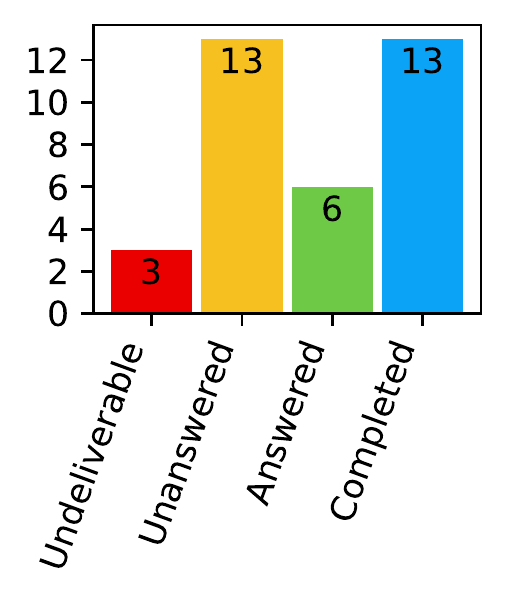}
        \caption{Status of correspondence}
        \label{fig:correspondence:status}
    \end{subfigure}
    \begin{subfigure}{.69\textwidth}
        \includegraphics[width=\textwidth]{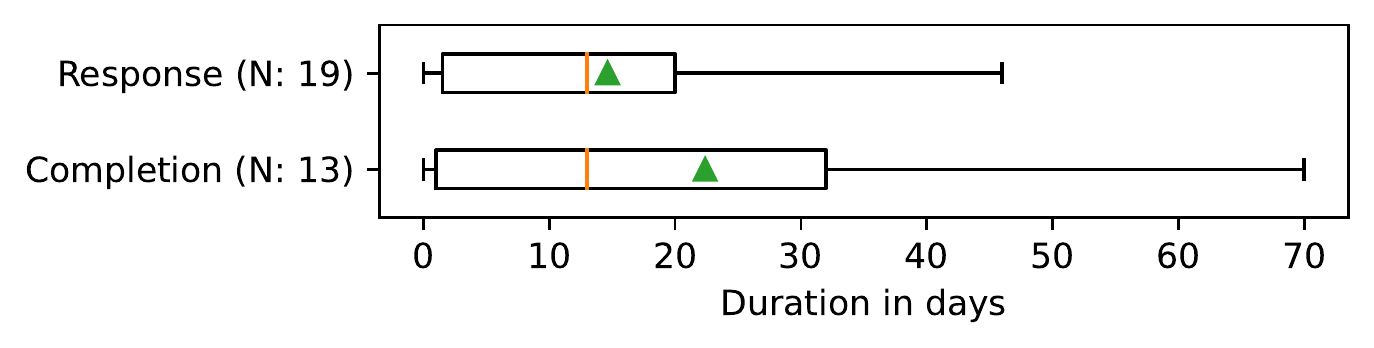}
        \caption{Time of correspondence}
        \label{fig:correspondence:duration}
        
        \includegraphics[width=\textwidth]{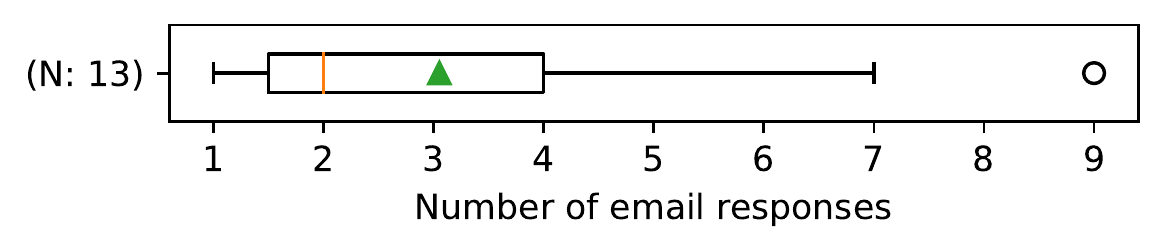}
        \caption{Frequency of correspondence}
        \label{fig:correspondence:frequency}
    \end{subfigure}
    \caption{Correspondence to the artifact recovery request}
    \label{fig:correspondence}
\end{figure}

Out of the 35 approached corresponding authors, 19 answered the recovery request, and 13 completed it. We could not reach three authors despite searching for a valid contact. The distribution of correspondence status is visualized in \Cref{fig:correspondence:status}. It took, on average, 14.6 days for a corresponding author to reply to our request and 22.4 additional days to complete the request. On average, a request was resolved in an exchange of 3 emails with the corresponding author. The distributions of these statistics are visualized in \Cref{fig:correspondence:duration} and \Cref{fig:correspondence:frequency}, respectively.

\subsection{Artifact Recovery Success}

The corresponding authors improved the availability of seven data sets (four of which follow open-access principles) and seven implementations (six following open-access principles). This increases the availability of data sets from 12.3\% (7/57, 1 open access) to 22.8\% (13/57, 5 open access) and the availability of implementations from 19.4\% (7/36, 0 open access) to 30.6\% (11/36, 6 open access). Authors further confirmed the unavailability of 21 data sets and six implementations and provided reasons for the inability to recover or disclose them.

\begin{figure}[h]
    \centering
    \begin{subfigure}{.48\textwidth}
        \centering
        \includegraphics[width=\textwidth]{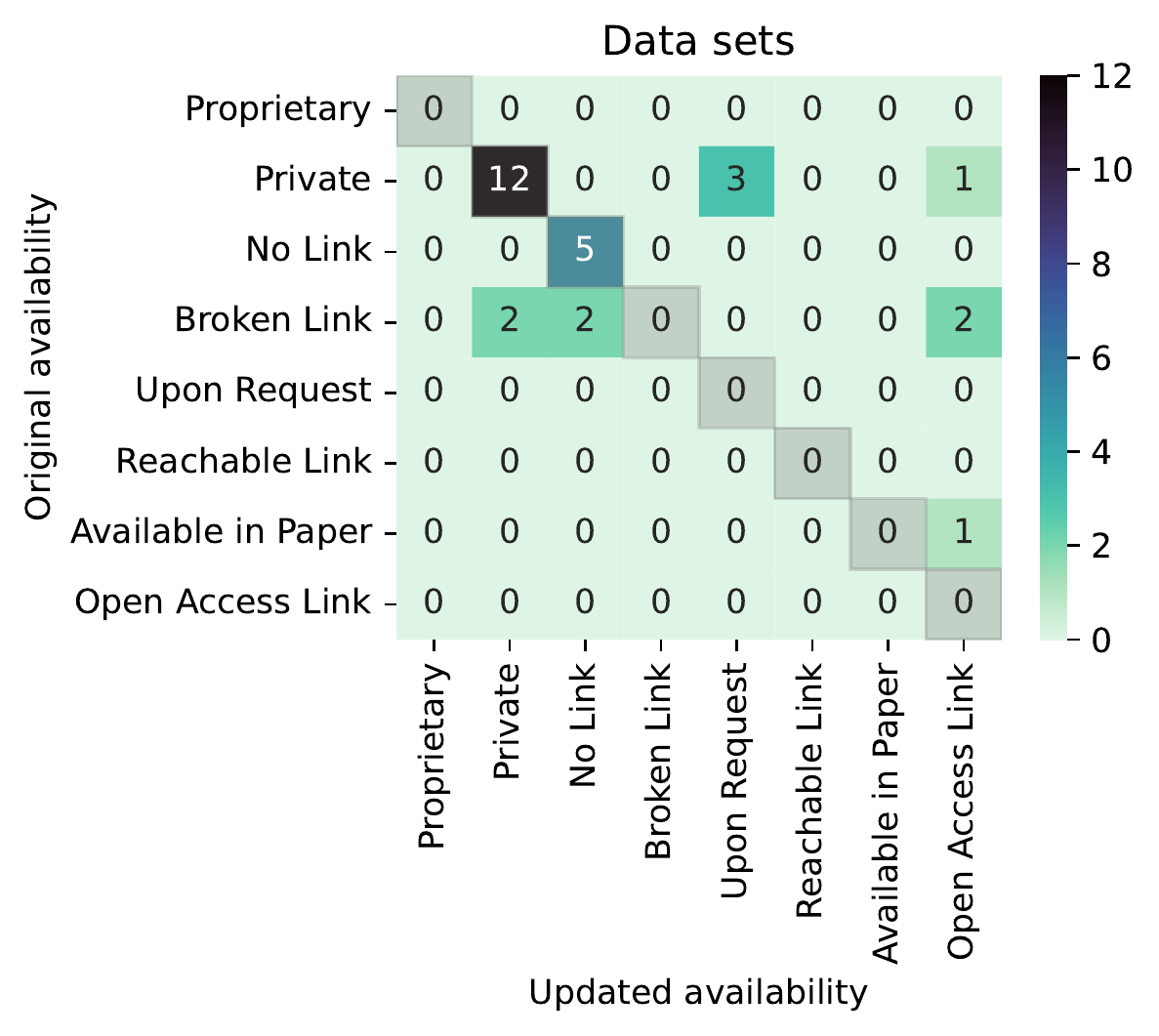}
        \caption{Change of availability in data sets}
        \label{fig:availability:datasets}
    \end{subfigure}
    \begin{subfigure}{.48\textwidth}
        \centering
        \includegraphics[width=\textwidth]{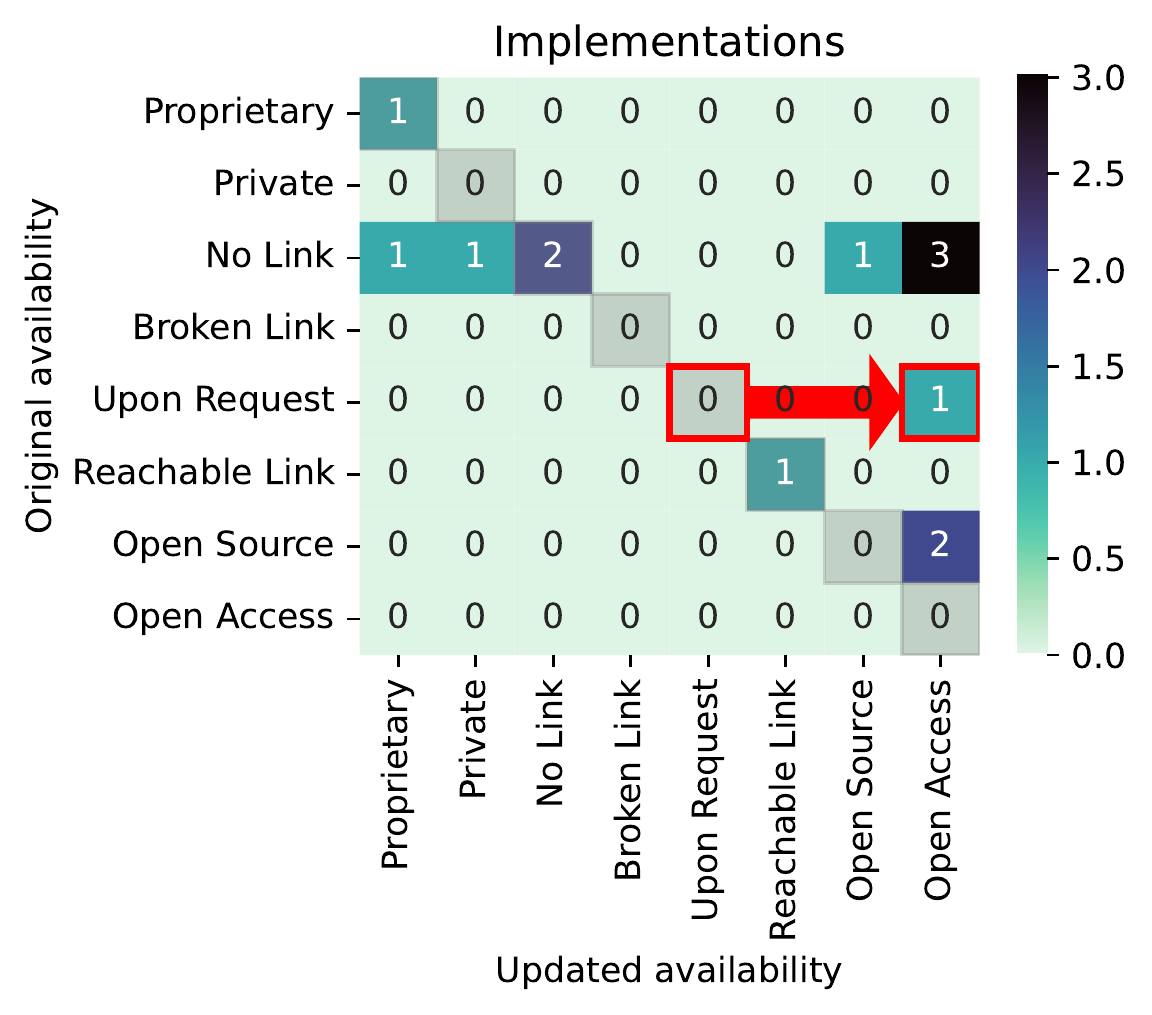}
        \caption{Change of availability in implementations}
        \label{fig:availability:approaches}
    \end{subfigure}
    
    \caption{Change of artifact availability}
    \label{fig:availability}
\end{figure}

\Cref{fig:availability} visualizes the success of the recovery request. The heatmap considers all artifacts (data sets in \Cref{fig:availability:datasets} and implementations in \Cref{fig:availability:approaches}) where the corresponding author completed the recovery request. The number in a cell represents the number of artifacts for which the original availability (on the y-axis) has been updated to the new availability (on the x-axis). The count of artifacts whose availability remained the same (e.g., because an author confirmed that the artifact could not be made more available) is reported on the diagonal (shaded gray). An improvement in the availability of an artifact contributes to cells to the right of the diagonal, a deterioration of the availability to the left.

For example, one implementation was previously available \textit{upon request}~\cite{li2016engineering}. Now that the authors disclosed the implementation following open access principles\footnote{Now publicly available at \url{https://doi.org/10.5281/zenodo.7484023}}, the entry moved three cells to the right (see \Cref{fig:availability:approaches}).

The inability to recover or disclose artifacts was reported as follows: among 21 unrecoverable data sets, 15 were lost (i.e., the author could not find them anymore or the contact, whom the author assumed had the data, was unreachable), and six could not be disclosed due to sensitive contents. Among the six unrecoverable implementations, three became proprietary, and three were lost.

\section{Discussion}
\label{sec:discussion}

Within the 70-day time frame for the process, authors of requirements quality publications recovered several data sets and implementations that are now available for reproduction of scientific results and reuse in future projects. We referenced the recovered artifacts in the requirements quality factor ontology~\cite{frattini2022live}\footnote{See \textit{Content} at \url{http://reqfactoront.com/}} as well as our replication package to make them accessible.

Additionally, the authors confirmed the unavailability of several more artifacts. While this does not actively improve the availability of artifacts for reproduction, it clarifies the ambiguous status of several data sets and implementations. Overall, when authors answered a recovery request, they either recovered their data or reported the inability to do so with helpful explanations. Recovery requests failed due to (1) no response, (2) the artifact being lost, or (3) the artifact containing sensitive information. We did not encounter other reasons for the failure of a recovery request, which corroborates the goodwill of the sampled requirements quality community in its commitment to open science. This stands in contrast to the experience of other artifact recovery attempts, where researchers encountered reasons like requests for reimbursements or not seeing any personal gain in the recovery~\cite{gabelica2022many}.

We cannot claim that our observations are universally valid for the software (requirements) engineering community due to the limitations of our study. For one, the set of primary studies was obtained via convenience sampling from a previous study~\cite{frattini2022live}. This sample has known limitations as several primary studies relevant to requirements quality literature are missing. Hence, the results of recovery success and correspondence do not represent the complete requirements quality literature and research community. Furthermore, our conclusion regarding the status of correspondence, especially the status of \textit{unanswered} and \textit{answered} requests, is limited by how we decided to approach corresponding authors. Using emails as the mean of communication impedes the response rate since they are often abandoned with a change of affiliation~\cite{wren2006mail}. The limited success of correspondence is a consequence of the time frame and communication channels used in this study rather than an indicator of the research community's attitude towards open science.

\section{Conclusion}
\label{sec:conclusion}

Both the credibility and reusability of previous publications in the requirements quality literature have been impeded by the unavailability of data sets and implementations. We requested corresponding authors of 57 publications to disclose their artifacts according to the open science principles. We improved the availability of seven data sets and seven implementations, several of which now follow open science principles.

With this study, we want to raise awareness about the importance of recovering artifacts associated with older publications. While adherence to the open science principles recently rose thanks to comprehensive guidelines (see \cite{mendez2020open}) or community initiatives such as artifact evaluation tracks at conferences, they are rarely applied retroactively to previous publications. Furthermore, we hope that the material we created will support researchers in areas that heavily rely on artifacts, such as NLP4RE, to recover more of them.

Our agenda, in the scope of the requirements quality factor ontology\footnote{\url{http://reqfactoront.com}}, includes providing a central repository of updated information on the availability and location of relevant artifacts. We invite researchers to contribute to this cause and strengthen the evidence in our field.

\begin{acknowledgments}
  The KKS foundation supported this work through the S.E.R.T. Research Profile project at Blekinge Institute of Technology. We additionally thank the reviewers for their valuable feedback upon which the manuscript was improved.
\end{acknowledgments}

\bibliography{references}

\begin{thebibliography}{17}
\expandafter\ifx\csname natexlab\endcsname\relax\def\natexlab#1{#1}\fi
\providecommand{\url}[1]{\texttt{#1}}
\providecommand{\href}[2]{#2}
\providecommand{\path}[1]{#1}
\providecommand{\DOIprefix}{doi:}
\providecommand{\ArXivprefix}{arXiv:}
\providecommand{\URLprefix}{URL: }
\providecommand{\Pubmedprefix}{pmid:}
\providecommand{\doi}[1]{\href{http://dx.doi.org/#1}{\path{#1}}}
\providecommand{\Pubmed}[1]{\href{pmid:#1}{\path{#1}}}
\providecommand{\bibinfo}[2]{#2}
\ifx\xfnm\relax \def\xfnm[#1]{\unskip,\space#1}\fi
\bibitem[{Montgomery et~al.(2022)Montgomery, Fucci, Bouraffa, Scholz, and
  Maalej}]{montgomery2022empirical}
\bibinfo{author}{L.~Montgomery}, \bibinfo{author}{D.~Fucci},
  \bibinfo{author}{A.~Bouraffa}, \bibinfo{author}{L.~Scholz},
  \bibinfo{author}{W.~Maalej},
\newblock \bibinfo{title}{Empirical research on requirements quality: a
  systematic mapping study},
\newblock \bibinfo{journal}{Requirements Engineering}  (\bibinfo{year}{2022})
  \bibinfo{pages}{1--27}.
\bibitem[{Frattini et~al.(2022)Frattini, Montgomery, Fischbach,
  Unterkalmsteiner, Mendez, and Fucci}]{frattini2022live}
\bibinfo{author}{J.~Frattini}, \bibinfo{author}{L.~Montgomery},
  \bibinfo{author}{J.~Fischbach}, \bibinfo{author}{M.~Unterkalmsteiner},
  \bibinfo{author}{D.~Mendez}, \bibinfo{author}{D.~Fucci},
\newblock \bibinfo{title}{A live extensible ontology of quality factors for
  textual requirements},
\newblock in: \bibinfo{booktitle}{2022 IEEE 30th International Requirements
  Engineering Conference (RE)}, \bibinfo{organization}{IEEE},
  \bibinfo{year}{2022}, pp. \bibinfo{pages}{274--280}.
\bibitem[{Mendez et~al.(2020)Mendez, Graziotin, Wagner, and
  Seibold}]{mendez2020open}
\bibinfo{author}{D.~Mendez}, \bibinfo{author}{D.~Graziotin},
  \bibinfo{author}{S.~Wagner}, \bibinfo{author}{H.~Seibold},
\newblock \bibinfo{title}{Open science in software engineering},
\newblock in: \bibinfo{booktitle}{Contemporary empirical methods in software
  engineering}, \bibinfo{publisher}{Springer}, \bibinfo{year}{2020}, pp.
  \bibinfo{pages}{477--501}.
\bibitem[{Anda et~al.(2008)Anda, Sj{\o}berg, and Mockus}]{anda2008variability}
\bibinfo{author}{B.~C. Anda}, \bibinfo{author}{D.~I. Sj{\o}berg},
  \bibinfo{author}{A.~Mockus},
\newblock \bibinfo{title}{Variability and reproducibility in software
  engineering: A study of four companies that developed the same system},
\newblock \bibinfo{journal}{TSE} \bibinfo{volume}{35} (\bibinfo{year}{2008})
  \bibinfo{pages}{407--429}.
\bibitem[{Tennant et~al.(2019)Tennant, Beamer, Bosman, Brembs, Chung, Clement,
  Crick, Dugan, Dunning et~al.}]{tennant2019foundations}
\bibinfo{author}{J.~Tennant}, \bibinfo{author}{J.~Beamer},
  \bibinfo{author}{J.~Bosman}, \bibinfo{author}{B.~Brembs},
  \bibinfo{author}{N.~C. Chung}, \bibinfo{author}{G.~Clement},
  \bibinfo{author}{T.~Crick}, \bibinfo{author}{J.~Dugan},
  \bibinfo{author}{A.~Dunning}, et~al.,
\newblock \bibinfo{title}{Foundations for open scholarship strategy
  development}  (\bibinfo{year}{2019}).
\bibitem[{Kidwell et~al.(2016)Kidwell, Lazarevi{\'c}, Baranski, Hardwicke,
  Piechowski, Falkenberg, Kennett, Slowik et~al.}]{kidwell2016badges}
\bibinfo{author}{M.~C. Kidwell}, \bibinfo{author}{L.~B. Lazarevi{\'c}},
  \bibinfo{author}{E.~Baranski}, \bibinfo{author}{T.~E. Hardwicke},
  \bibinfo{author}{S.~Piechowski}, \bibinfo{author}{L.-S. Falkenberg},
  \bibinfo{author}{C.~Kennett}, \bibinfo{author}{A.~Slowik}, et~al.,
\newblock \bibinfo{title}{Badges to acknowledge open practices: A simple,
  low-cost, effective method for increasing transparency},
\newblock \bibinfo{journal}{PLoS biology} \bibinfo{volume}{14}
  (\bibinfo{year}{2016}) \bibinfo{pages}{e1002456}.
\bibitem[{Nosek et~al.(2018)Nosek, Ebersole, DeHaven, and
  Mellor}]{nosek2018preregistration}
\bibinfo{author}{B.~A. Nosek}, \bibinfo{author}{C.~R. Ebersole},
  \bibinfo{author}{A.~C. DeHaven}, \bibinfo{author}{D.~T. Mellor},
\newblock \bibinfo{title}{The preregistration revolution},
\newblock \bibinfo{journal}{Proceedings of the National Academy of Sciences}
  \bibinfo{volume}{115} (\bibinfo{year}{2018}) \bibinfo{pages}{2600--2606}.
\bibitem[{Gabelica et~al.(2022)Gabelica, Boj{\v{c}}i{\'c}, and
  Puljak}]{gabelica2022many}
\bibinfo{author}{M.~Gabelica}, \bibinfo{author}{R.~Boj{\v{c}}i{\'c}},
  \bibinfo{author}{L.~Puljak},
\newblock \bibinfo{title}{Many researchers were not compliant with their
  published data sharing statement: mixed-methods study},
\newblock \bibinfo{journal}{Journal of Clinical Epidemiology}
  (\bibinfo{year}{2022}).
\bibitem[{Wagner et~al.(2019)Wagner, Fern{\'a}ndez, Felderer, Vetr{\`o},
  Kalinowski, Wieringa, Pfahl, Conte, Christiansson, Greer
  et~al.}]{wagner2019status}
\bibinfo{author}{S.~Wagner}, \bibinfo{author}{D.~M. Fern{\'a}ndez},
  \bibinfo{author}{M.~Felderer}, \bibinfo{author}{A.~Vetr{\`o}},
  \bibinfo{author}{M.~Kalinowski}, \bibinfo{author}{R.~Wieringa},
  \bibinfo{author}{D.~Pfahl}, \bibinfo{author}{T.~Conte},
  \bibinfo{author}{M.-T. Christiansson}, \bibinfo{author}{D.~Greer}, et~al.,
\newblock \bibinfo{title}{Status quo in requirements engineering: A theory and
  a global family of surveys},
\newblock \bibinfo{journal}{TOSEM} \bibinfo{volume}{28} (\bibinfo{year}{2019})
  \bibinfo{pages}{1--48}.
\bibitem[{Mendez and Wagner(2013)}]{fernandez2013naming}
\bibinfo{author}{D.~Mendez}, \bibinfo{author}{S.~Wagner},
\newblock \bibinfo{title}{Naming the pain in requirements engineering: Design
  of a global family of surveys and first results from germany},
\newblock in: \bibinfo{booktitle}{Proceedings of the 17th International
  Conference on Evaluation and Assessment in Software Engineering},
  \bibinfo{year}{2013}, pp. \bibinfo{pages}{183--194}.
\bibitem[{Ezzini et~al.(2021)Ezzini, Abualhaija, Arora, Sabetzadeh, and
  Briand}]{ezzini2021using}
\bibinfo{author}{S.~Ezzini}, \bibinfo{author}{S.~Abualhaija},
  \bibinfo{author}{C.~Arora}, \bibinfo{author}{M.~Sabetzadeh},
  \bibinfo{author}{L.~C. Briand},
\newblock \bibinfo{title}{Using domain-specific corpora for improved handling
  of ambiguity in requirements},
\newblock in: \bibinfo{booktitle}{2021 IEEE/ACM 43rd International Conference
  on Software Engineering (ICSE)}, \bibinfo{organization}{IEEE},
  \bibinfo{year}{2021}, pp. \bibinfo{pages}{1485--1497}.
\bibitem[{Zhao et~al.(2021)Zhao, Alhoshan, Ferrari, Letsholo, Ajagbe, Chioasca,
  and Batista-Navarro}]{zhao2021natural}
\bibinfo{author}{L.~Zhao}, \bibinfo{author}{W.~Alhoshan},
  \bibinfo{author}{A.~Ferrari}, \bibinfo{author}{K.~J. Letsholo},
  \bibinfo{author}{M.~A. Ajagbe}, \bibinfo{author}{E.-V. Chioasca},
  \bibinfo{author}{R.~T. Batista-Navarro},
\newblock \bibinfo{title}{Natural language processing for requirements
  engineering: A systematic mapping study},
\newblock \bibinfo{journal}{ACM Computing Surveys (CSUR)} \bibinfo{volume}{54}
  (\bibinfo{year}{2021}) \bibinfo{pages}{1--41}.
\bibitem[{Dalpiaz et~al.(2018)Dalpiaz, Ferrari, Franch, and
  Palomares}]{dalpiaz2018natural}
\bibinfo{author}{F.~Dalpiaz}, \bibinfo{author}{A.~Ferrari},
  \bibinfo{author}{X.~Franch}, \bibinfo{author}{C.~Palomares},
\newblock \bibinfo{title}{Natural language processing for requirements
  engineering: The best is yet to come},
\newblock \bibinfo{journal}{IEEE software} \bibinfo{volume}{35}
  (\bibinfo{year}{2018}) \bibinfo{pages}{115--119}.
\bibitem[{Minocher et~al.(2020)Minocher, Atmaca, Bavero, McElreath, and
  Beheim}]{minocher2020reproducibility}
\bibinfo{author}{R.~Minocher}, \bibinfo{author}{S.~Atmaca},
  \bibinfo{author}{C.~Bavero}, \bibinfo{author}{R.~McElreath},
  \bibinfo{author}{B.~Beheim},
\newblock \bibinfo{title}{Reproducibility improves exponentially over 63 years
  of social learning research}  (\bibinfo{year}{2020}).
\bibitem[{Baltes and Ralph(2022)}]{baltes2022sampling}
\bibinfo{author}{S.~Baltes}, \bibinfo{author}{P.~Ralph},
\newblock \bibinfo{title}{Sampling in software engineering research: A critical
  review and guidelines},
\newblock \bibinfo{journal}{Empirical Software Engineering}
  \bibinfo{volume}{27} (\bibinfo{year}{2022}) \bibinfo{pages}{1--31}.
\bibitem[{Li et~al.(2016)Li, Horkoff, Liu, Borgida, Guizzardi, and
  Mylopoulos}]{li2016engineering}
\bibinfo{author}{F.-L. Li}, \bibinfo{author}{J.~Horkoff},
  \bibinfo{author}{L.~Liu}, \bibinfo{author}{A.~Borgida},
  \bibinfo{author}{G.~Guizzardi}, \bibinfo{author}{J.~Mylopoulos},
\newblock \bibinfo{title}{Engineering requirements with desiree: An empirical
  evaluation},
\newblock in: \bibinfo{booktitle}{International Conference on Advanced
  Information Systems Engineering}, \bibinfo{organization}{Springer},
  \bibinfo{year}{2016}, pp. \bibinfo{pages}{221--238}.
\bibitem[{Wren et~al.(2006)Wren, Grissom, and Conway}]{wren2006mail}
\bibinfo{author}{J.~D. Wren}, \bibinfo{author}{J.~E. Grissom},
  \bibinfo{author}{T.~Conway},
\newblock \bibinfo{title}{E-mail decay rates among corresponding authors in
  medline: The ability to communicate with and request materials from authors
  is being eroded by the expiration of e-mail addresses},
\newblock \bibinfo{journal}{EMBO reports} \bibinfo{volume}{7}
  (\bibinfo{year}{2006}) \bibinfo{pages}{122--127}.

\end{thebibliography}
\end{document}